\documentclass[reprint,amsmath,amssymb,aps,prl,superscriptaddress,showpacs,citeautoscript,floatfix]{revtex4-1}
\usepackage{graphicx}
\usepackage{color}
\usepackage{dcolumn}
\usepackage{bm}
\usepackage{amsmath}
\usepackage{amssymb}
\usepackage[]{units}
\usepackage{booktabs}
\DeclareGraphicsExtensions{.eps,.png,.jpg,.pdf}
\graphicspath{{Figures/}}

\begin{document}

\author{M. Wolloch}
\affiliation{Department of Physics, Informatics and Mathematics, University of Modena and Reggio Emilia, Via Campi, 213/A 41125 Modena, Italy}
\author{G. Levita}
\affiliation{CNR-Institute of Nanoscience, S3 Center, Via Campi 213/A, 41125 Modena, Italy}
\author{P. Restuccia}
\affiliation{Department of Physics, Informatics and Mathematics, University of Modena and Reggio Emilia, Via Campi, 213/A 41125 Modena, Italy}
\author{M. C. Righi}
\email{mcrighi@unimore.it}
\affiliation{Department of Physics, Informatics and Mathematics, University of Modena and Reggio Emilia, Via Campi, 213/A 41125 Modena, Italy}
\affiliation{CNR-Institute of Nanoscience, S3 Center, Via Campi 213/A, 41125 Modena, Italy}
\email{mcrighi@unimore.it}

\title[Charge density, adhesion, and friction]
{Interfacial charge density and its connection to adhesion and frictional forces}

\begin{abstract}
We derive a connection between the intrinsic tribological properties and the electronic properties of a solid interface. 
In particular, we show that the adhesion and frictional forces
are dictated by the electronic charge redistribution 
occurring due to the relative displacements of the two surfaces in contact.
We define a figure of merit to quantify such charge redistribution and show 
that simple functional relations hold for a wide series of interactions including metallic, covalent and physical bonds.
This suggests unconventional ways of measuring friction 
by recording the evolution of the interfacial electronic charge during sliding. 
Finally, we explain that the key mechanism to reduce adhesive friction is to inhibit the charge flow at the interface and provide examples of this mechanism in common lubricant additives. 

\end{abstract}

\maketitle

Friction and adhesion are common phenomena that impact many fields from nanotechnologies to earthquakes, but their fundamental origin is still largely unknown~\cite{krylov:14}. The reason resides in the fact that even for macroscopic objects, friction and adhesion are governed by microscopic contacts, whererather uniformly (black solid curve of Fig.~\ref{fig:totCharge}(d)), while for the on-top configuration significantly less charge (red dashed curve of Fig.~\ref{fig:totCharge}(d)) is concentrated at the interface and it is less homogeneous Fig.~\ref{fig:Corrugation}(b) and (d). the atomistic interactions of quantum mechanical origin ultimately determine the tribological response~\cite{jacobs:17}. Thus, it is of great practical and theoretical importance to understand the connection between the electronic structure and the mechanical tribological properties of interfaces.
At the atomic level, adhesion is dictated by the chemical interaction between the surfaces in contact and adhesive friction arises because this interaction changes as a function of the relative lateral position of the two surfaces. In turn, the adhesion and frictional forces can be understood by analyzing the charge density $\rho$ in the region of the interface, and more specifically the charge redistribution occurring when the two surfaces are moved relative to each other, either to initially form the interface or during sliding.

Understanding the connection between the interfacial charge density and adhesive friction is of paramount importance to design lubricant additives~\cite{zhou:17,xiao:17}. However, nowadays most of the research on lubricants is conducted empirically due to the lack of predictive understanding, which we believe can be achieved by the analysis of the electronic interfacial properties. The functionality of some solid and boundary lubricants is, in fact, based on their capability of decreasing the adhesive interactions between the surfaces in contact.
In this Letter we show that this functionality relies precisely on their ability to reduce the charge density at the interface.

The advent of scanning probe techniques in tribology, such as friction force microscopy, has allowed scientists to obtain friction maps between nanometer-size contacts with nano- and piconewton resolution~\cite{szlufarska:08}. Here we show that the friction maps directly reflect the charge density maps recorded during sliding. Therefore simultaneous measurement of the tribological and electronic interfacial properties should be attempted.

One of the most important figures of merit in tribology is the work of separation, which corresponds to the energy required to separate two surfaces from contact and is the opposite of the adhesion energy $W_\mathrm{sep}=-E_\mathrm{adh}$. The variation of $E_\mathrm{adh}$ as a function of the lateral displacement during sliding is what causes the appearance of frictional forces. This variation is described by a potential energy surface (PES), $V(x,y,z_\mathrm{eq})$, which in the dislocation community is known as the $\gamma$-surface, and where $z_\mathrm{eq}$ is the surface separation at zero load. The absolute minimum of the PES corresponds to the adhesion energy, $E_\mathrm{adh}=V(x_\mathrm{eq},y_\mathrm{eq},z_\mathrm{eq})$, while the energy difference between the minimum and the maximum of the PES is referred to as the corrugation and will be denoted by $\Delta V$ in the following. This number is especially important since it is equivalent to the maximum amount of energy per unit area that might be dissipated by frictional processes.

Historically simple sinusoidal energy profiles have played a significant role in describing the elementary mechanisms of friction, where stick-slip (or continuous motion) from minimum to minimum is analyzed~\cite{prandtl:28,tomlinson:29}. Later, ab-initio data were used to generate these energy profiles~\cite{zhong:90,tomanek:91}, and most recently, the whole two dimensional PES has been used to analyze friction~\cite{righi:07,zilibotti:09,zilibotti:11,levita:14,cahangirov:12,wolloch:14a}.
Using the whole PES allows one to identify friction anisotropy and the easiest sliding path (or minimum energy path) which carries the highest statistical weight.

In 2012, Reguzzoni and coworkers used the interfacial charge density, especially charge density difference profiles, to gain insight into the frictional characteristics of graphene sliding on graphene~\cite{reguzzoni:12}. In recent years other publications built on this idea~\cite{cahangirov:13,levita:14,restuccia:16,wang:17}.

In this Letter we use density functional theory (DFT) to present our discovery of a deeper connection between adhesion, the PES, and interfacial charge density variations. We consider a large number of solids and find a linear relation between the amount of charge that is redistributed during the formation of an interface, and the adhesion energy. Moreover, a simple functional relation between the strength of adhesion $E_\mathrm{adh}$ and the corrugation of the PES $\Delta V$ is discovered. 
We also explain that one key function of lubricant materials is surface passivation to impede the charge flow at the interface~\cite{konicek:08,zilibotti:13,barros-bouchet:15,marchetto:17}. Finally we show that the PES corrugation and in turn adhesive friction are determined by the variation of the total charge density at the interface during sliding. An experimental verification of this observation using scanning probe techniques is proposed.

For all calculations we used the plane wave DFT package Quantum ESPRESSO~\cite{espresso}, the computational details can be found in the Supplementary Material (SM)~\cite{SM_Charge}.

\begin{figure}[thb]
  \includegraphics[width=\linewidth]{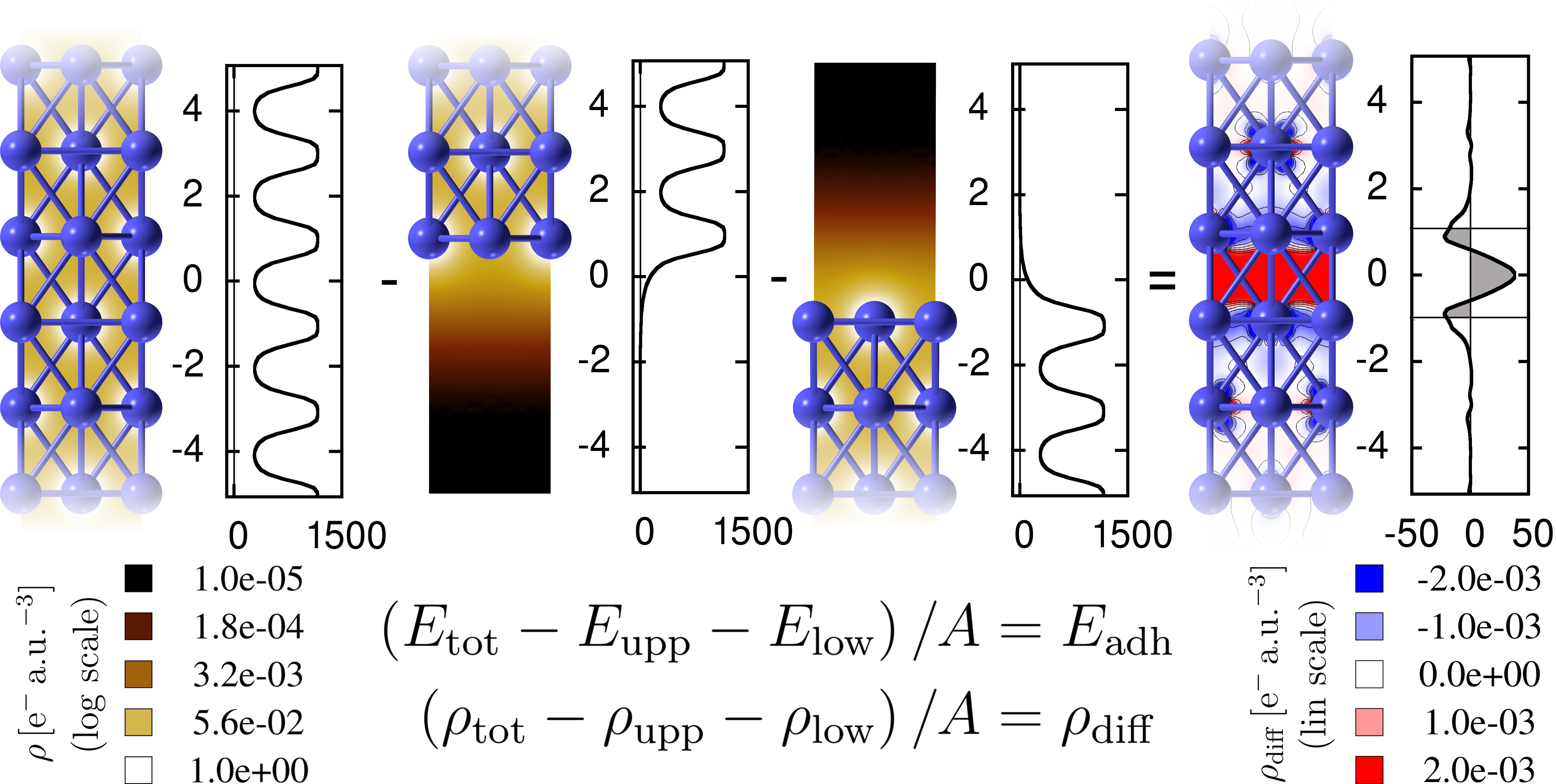}
  \caption{Calculation of $\rho_\mathrm{diff}$ using Fe(110) surfaces as an example. The total charge density $\rho$ is visualized in a black-brown-white color scheme on a log10 scale. $\rho_\mathrm{diff}$ is visualized in a color scheme from blue (depletion) to red (accumulation) of charge. Line plots are planar averages of $\rho_\mathrm{diff}$ and $\rho$ respectively on a scale of 10$^{-3}$ electrons per \AA$^{3}$.}
  \label{fig:Adhesion}
\end{figure}

Fig.~\ref{fig:Adhesion} shows how the charge displacement, or charge density difference $\rho_\mathrm{diff}$, is calculated for an example system of Fe(110) surfaces. The procedure is exactly the same for the adhesion energy $E_\mathrm{adh}$, as is indicated by the formulas in Fig.~\ref{fig:Adhesion}: The charge density (energy) of the two parts is subtracted from the charge density (energy) of the combined interface. Total charge is plotted using a black-brown-white color scale and $\rho_\mathrm{diff}$ is shown on a blue-white-red scale to visualize depletion (blue) and accumulation (red).
Planar averages are also shown as a line profile in the direction normal to the interface allowing one to quickly see where most charge is redistributed upon interface formation.
There the units are 10$^{-3}$ electrons per \AA$^{3}$, since we divide by the surface area of the simulation cell to compare rather uniformly (black solid curve of Fig.~\ref{fig:totCharge}(d)), while for the on-top configuration significantly less charge (red dashed curve of Fig.~\ref{fig:totCharge}(d)) is concentrated at the interface and it is less homogeneous Fig.~\ref{fig:Corrugation}(b) and (d).different systems.
It is immediately clear that the charge density of the separated slabs decays exponentially in the vacuum. If the slabs are brought together, the charge near the surfaces is depleted slightly and accumulated in the interface region, which we define as the space between the lowest atomic layer of the top slab (at $z_0$) and the highest one on the bottom slab (at $-z_0$).

To quantify this redistribution of charge density we integrate the absolute value of the profile of $\rho_\mathrm{diff}$ in the interface region, and normalize with respect to its width. We call this figure of merit, which measures both depletion and accumulation of charge density within the interface $\rho_\mathrm{redist}$,rather uniformly (black solid curve of Fig.~\ref{fig:totCharge}(d)), while for the on-top configuration significantly less charge (red dashed curve of Fig.~\ref{fig:totCharge}(d)) is concentrated at the interface and it is less homogeneous Fig.~\ref{fig:Corrugation}(b) and (d).
\begin{equation}
\rho_\mathrm{redist}=\frac{1}{2z_0}\int^{z_0}_{-z_0}\!\!\!|\rho_\mathrm{diff}| dz \quad .
\label{eq:F}
\end{equation}
This redistribution is generally more important than the (usually very small) net flow of charge into the interface region. $\rho_\mathrm{redist}$ corresponds to the shaded area in the line profile of Fig.~\ref{fig:Adhesion}, normalized to the interface width $2z_0$. Because of this normalization, which is necessary to take into account the effects of atom volumes and bond lengths, $\rho_\mathrm{redist}$ has the unit of a charge density. More details on its properties can be found in the SM~\cite{SM_Charge}.

\begin{figure}[hbt]
  \includegraphics[width=1.0\linewidth]{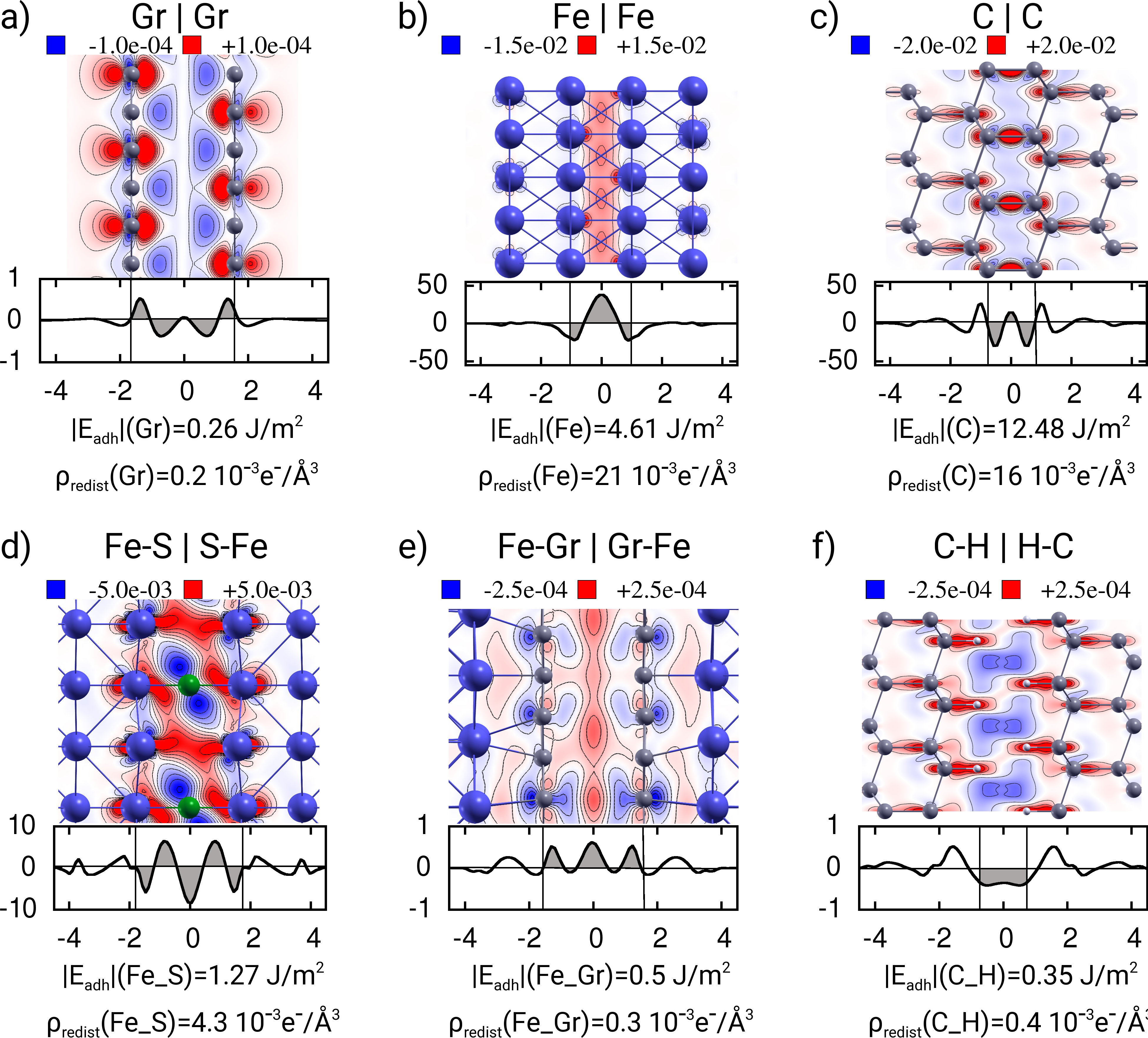}
  \caption{Charge density differences $\rho_\mathrm{diff}$ of different materials and their connection to adhesion values. Scales differ from plot to plot. The effects of partial and full passivation of Fe(110) and Diamond(111) on $\rho_\mathrm{diff}$ and $W_\mathrm{sep}$ are shown. $\rho_\mathrm{redist}$ values are shown as well and indicated by grey shaded areas. Units are the same as in Fig.~\ref{fig:Adhesion}.}
  \label{fig:ChargeDiff}
\end{figure}

\begin{figure}[hbt]
  \includegraphics[width=1.0\linewidth]{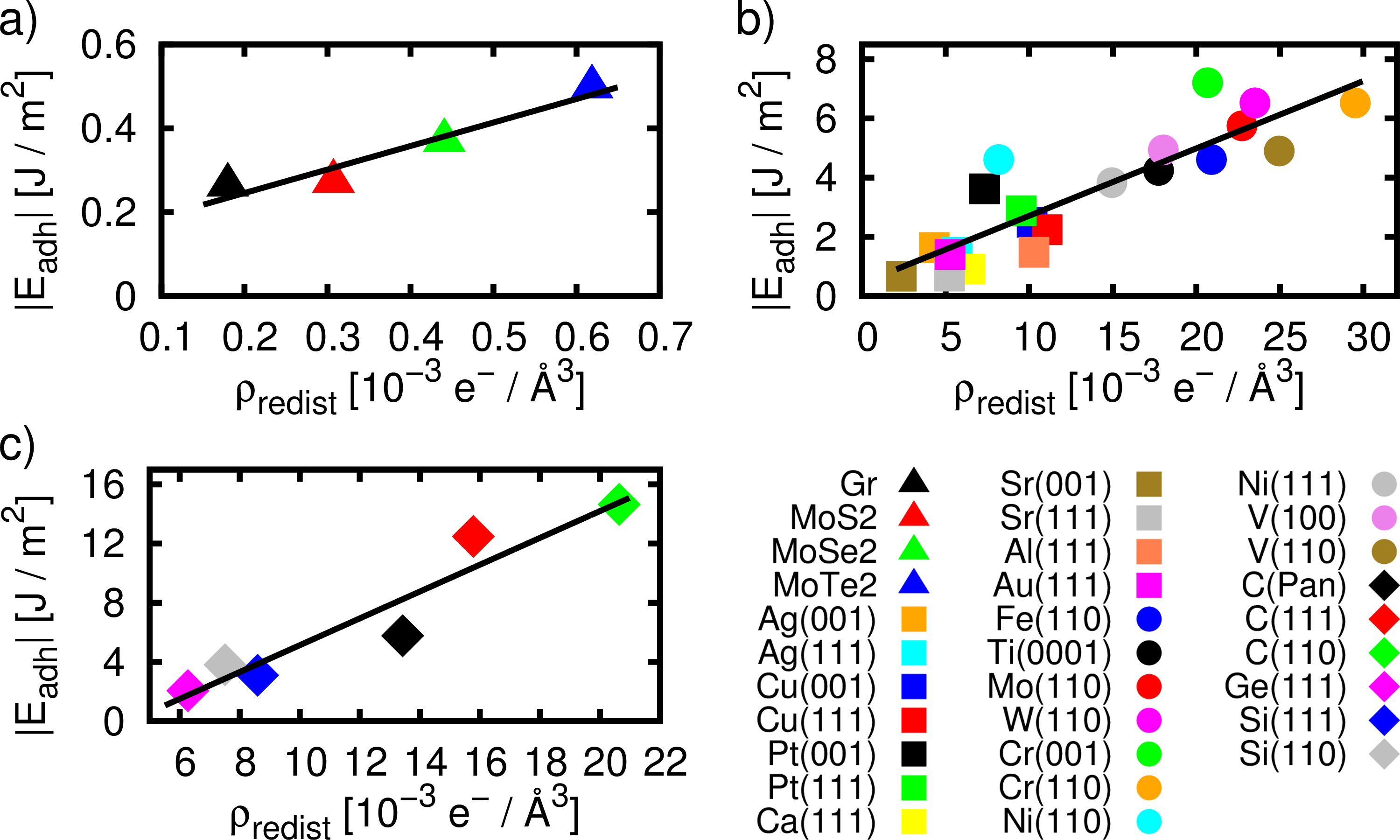}
  \caption{Adhesion energy versus charge redistribution $\rho_\mathrm{redist}$ for (a) vdW bonded materials, (b) simple and noble metals (squares) and non-noble transition metals (circles), and (c) covalently bonded materials. C(Pan) specifies the Pandey reconstruction.}
  \label{fig:LinearTrends}
\end{figure}

In the first row of Fig.~\ref{fig:ChargeDiff}, three materials with different bonding types are depicted: van der Waals (vdW) bonding for double-layer graphene~(a), abbreviated as Gr in all figures, metallic bonding for iron~(b), and covalent bonding for diamond~(c), abbreviated as C throughout the letter. The different scales in the color plots and the line profiles, as well as the different values of $E_\mathrm{adh}$, show a strong connection between $E_\mathrm{adh}$, $\rho_\mathrm{diff}$ and in turn $\rho_\mathrm{redist}$. The adhesion increases by one order of magnitude between graphene and Fe, and between Fe and diamond, and a large increase in the magnitude of $\rho_\mathrm{diff}$ can be observed as well. While the line profiles of $\rho_\mathrm{diff}$ are similar in magnitude for iron and diamond, the metallic system rearranges the charge much more uniformly at the interface center, while the charge is concentrated strongly along the directional carbon-carbon bonds in the insulator. While adhesion increases for C(111) compared to Fe(110), $\rho_\mathrm{redist}$ decreases. This indicates a different scaling of $\rho_\mathrm{redist}$ for different bonding types, which we will revisit in Fig.~\ref{fig:LinearTrends}.

Experiments have shown that lubricant additives containing sulfur or graphene can reduce the friction and wear of iron and steel significantly ~\cite{berman:13,righi:16,marchetto:17}. Likewise, friction at diamond interfaces is greatly diminished in the presence of H$_2$~\cite{barros-bouchet:12}. To investigate the root cause of these advantageous effects we investigate the influence of these species on $\rho_\mathrm{redist}$ and $E_\mathrm{adh}$ at Fe and C interfaces in the second row of Fig.~\ref{fig:ChargeDiff}.
The leftmost panel results from the addition of a $\nicefrac{1}{4}$ monolayer of sulfur at the iron surfaces, which is the most favorable coverage for Fe(110)~\cite{spencer:05}. This hinders charge accumulation at the interface compared to bare Fe, which in turn reduces $E_\mathrm{adh}$ by a factor of $\sim 3.5$.  The adhesion can be further reduced by higher coverage, as shown also in the case of phosphorus~\cite{barros-bouchet:15}. In panel (e) we see that the iron surfaces are fully passivated by chemisorped layers of graphene, which reduces the adhesion energy by one order of magnitude into the same range as double-layer graphene.
Finally in panel~(f) we present results for diamond (111) surfaces where the presence of hydrogen termination leads to full passivation. In this case no covalent bonds are formed between the surfaces and $\rho_\mathrm{diff}$ is reduced extremely at the interface center. The adhesion for hydrogenated diamond is nearly two orders of magnitude smaller than for bare surfaces. 
Comparing (b) with (d) and (e), as well as (c) with (f), in Fig.~\ref{fig:ChargeDiff}, allows for a clear understanding of the lubrication mechanism of these passivating species, which consists of preventing charge accumulation at the interface.

In Fig.~\ref{fig:LinearTrends} we correlate $\rho_\mathrm{redist}$ with the interfacial adhesion energy for a large set of different layered materials, metals and insulators/semiconductors.~\footnote{The planar averages of $\rho_\mathrm{diff}$ of all investigated materials can be found in the SM~\cite{SM_Charge}.}

It is remarkable that the redistribution of the charge at the interface is so directly related to the adhesion of such a wide variety of systems and surface terminations. The correlation within each bonding type is very good, with Pearson correlation coefficients for all fits are $\sim0.9$ or higher (see SM~\cite{SM_Charge} for more details). The different slopes of the linear fits allow to distinguish the bonding situation, something that usually is determined by analysis of $\nabla \rho$ and the Hessian matrix~\cite{Book_Bader:90,johnson:10}. For metals, Fig.~\ref{fig:LinearTrends}(b), we see a distinct grouping of noble and simple metals (squares) in the bottom left and the remaining transition metals (circles) on the top right. This is explained by the significant covalent bonding contribution of non-noble transition metals which increases both the $E_\mathrm{adh}$ and $\rho_\mathrm{redist}$.
It is important to note that for metals and layered materials the linear relation holds also very well if the independent variable $\rho_\mathrm{redist}$ is replaced by the height of the central peak of the planar average of $\rho_\mathrm{diff}$ (see Fig.~S1 in the SM~\cite{SM_Charge}).

\begin{figure}[hbt]
  \includegraphics[width=1.0\linewidth]{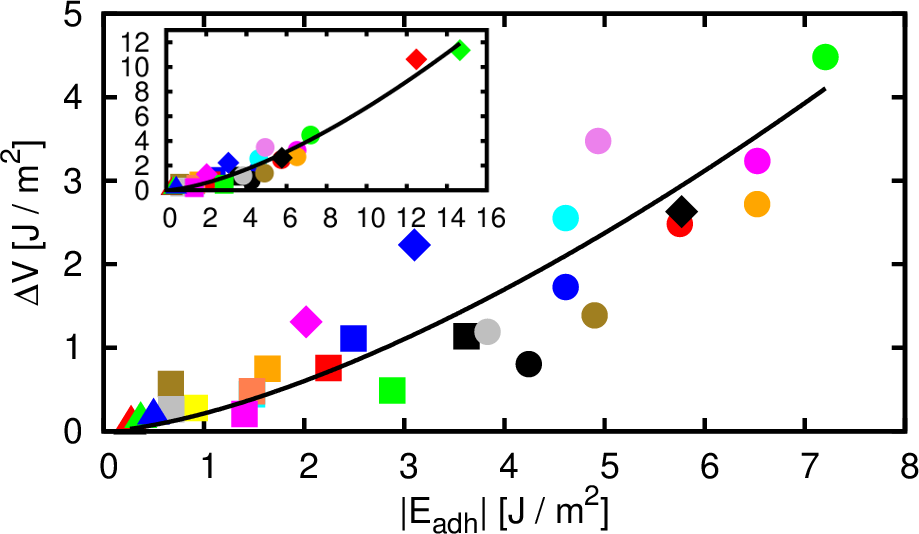}
  \caption{PES corrugation $\Delta V$ versus adhesion energy $|E_\mathrm{adh}|$. The main figure enlarges the region where most data are located while the inset shows all data. The black line fits all data with a power law and their Pearson correlation coefficient is $0.95$. Symbols and colors are the same as in Fig.~\ref{fig:LinearTrends}.}
  \label{fig:Corrugation_vs_Adhesion}
\end{figure}

\begin{figure}[hbt]
  \includegraphics[width=1.0\linewidth]{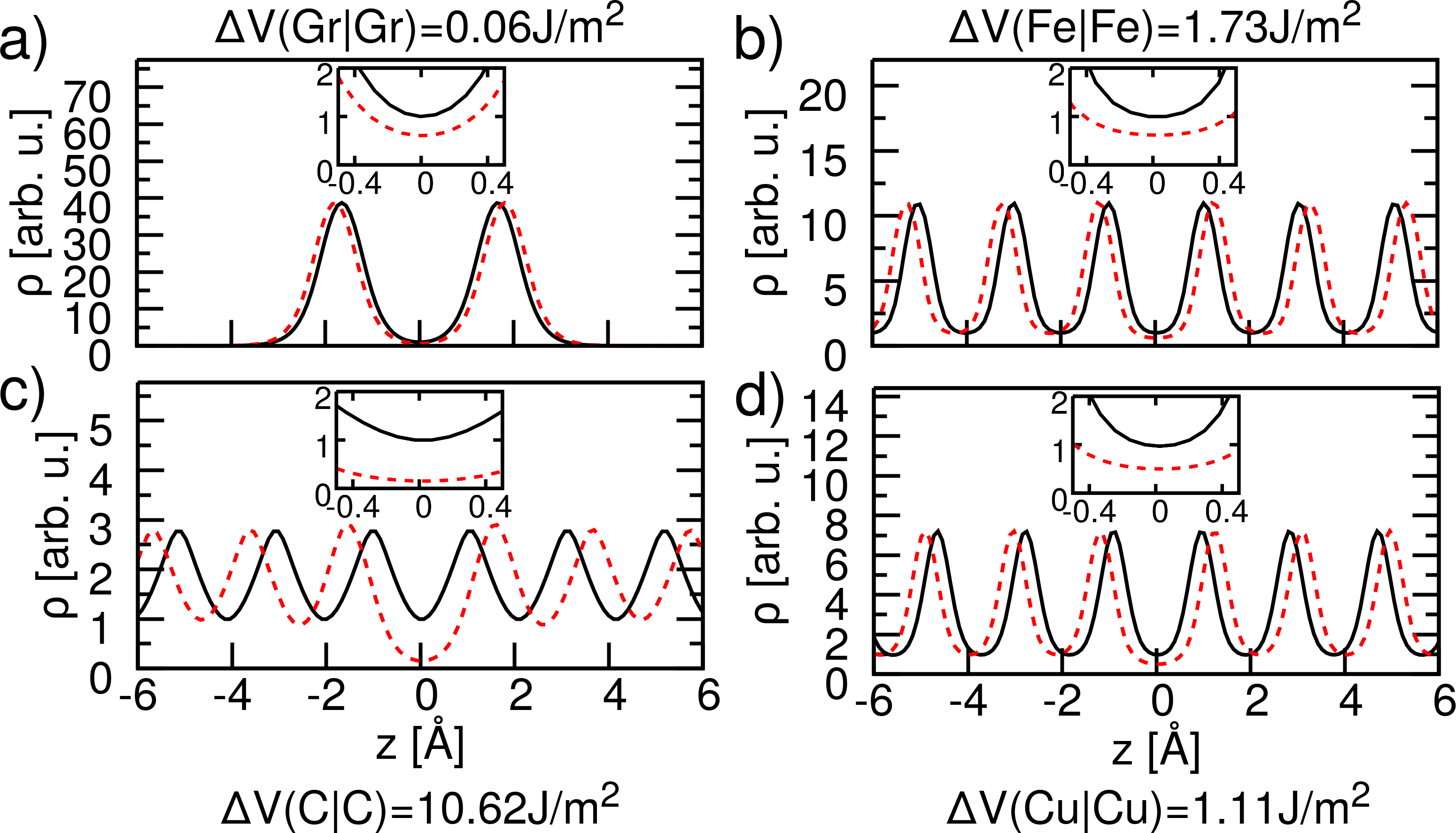}
  \caption{Total charge profiles for the interfaces of the materials presented in the first row of Fig.~\ref{fig:ChargeDiff} and Cu(001). Solid black lines are for the minimum configurations; dashed red lines are for the maxima. Charge is normalized to the respective value at the center of the interface in the minimum configuration.}
  \label{fig:totCharge}
\end{figure}

\begin{figure}[hbt]
  \includegraphics[width=1.0\linewidth]{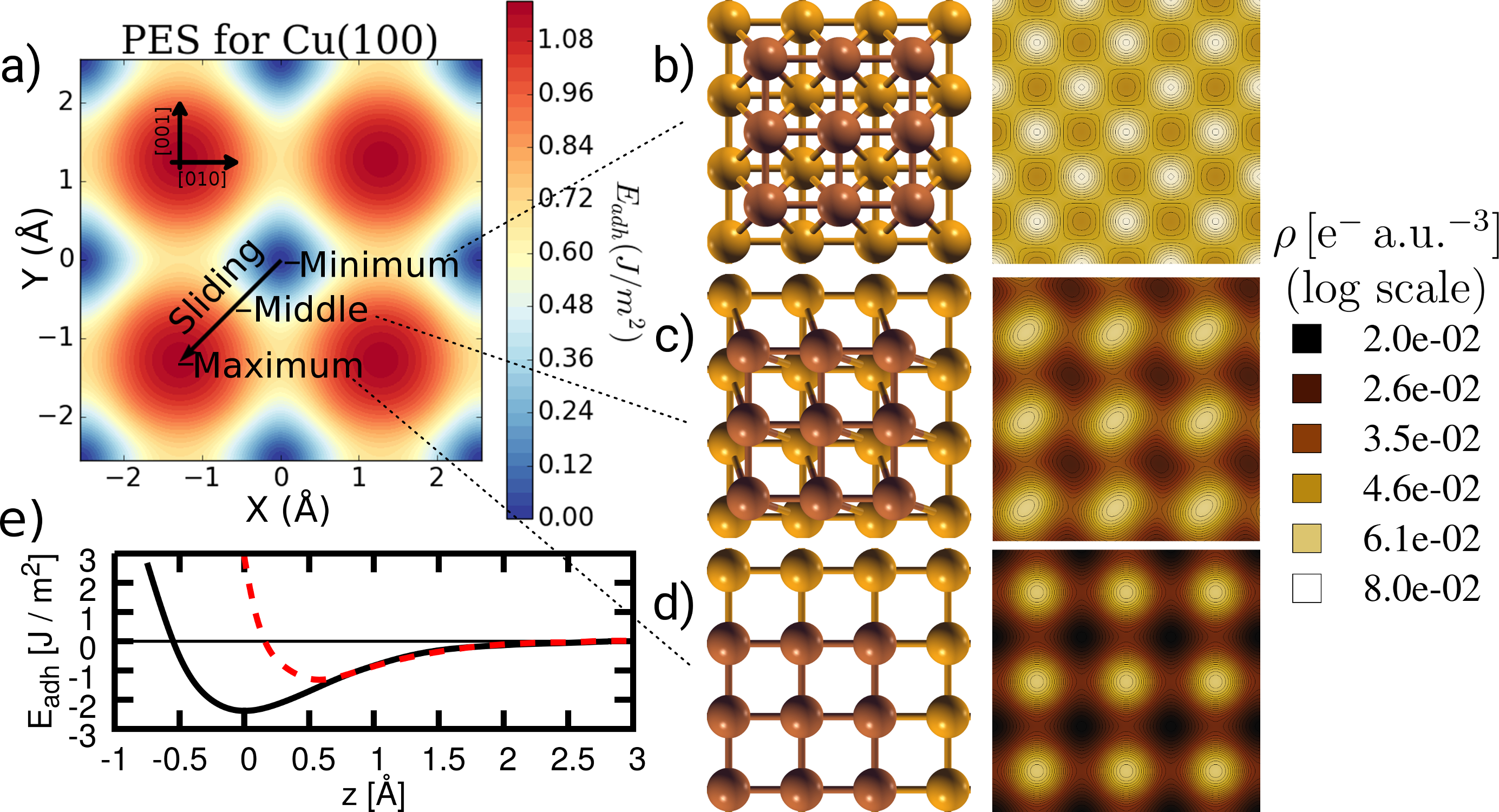}
  \caption{Potential energy surface of Cu(100) interface (a). Total charge densities at the interface for the minimum (b), a intermediate position (c), and the maximum (d), which are marked on the sliding path shown by the arrow in (a). Interaction potentials for minimum and maximum configurations are shown in (e).}
  \label{fig:Corrugation}
\end{figure}

In the following we analyze the potential corrugation $\Delta V$. First we consider the relation between $\Delta V$ and the adhesion energy $E_\mathrm{adh}$. As can be seen in Fig.~\ref{fig:Corrugation_vs_Adhesion}, we find great correlation independent from the bond type. All data now gather around a single curve, which can be very well fitted by a power law with exponent $\nicefrac{3}{2}$, ${\Delta V=aE_\mathrm{adh}^\frac{3}{2} }$ ($a$=\unit[0.21]{m J$^{-\frac{1}{2}}$}).
Close relations between adhesion and friction have been discussed before~\cite{yoshizawa:93,miyoshi:99,urbakh:04}, but a concrete functional relation has not been given yet. The added value of the power law is rather evident (although the exponent of $\nicefrac{3}{2}$ is not yet formally derived), as it permits to make precise predictions. The strong correlation between $E_\mathrm{adh}$ and $\Delta V$ is further evidence that reducing adhesion by surface passivation using lubricant additives leads to reduced friction (see Fig.~\ref{fig:ChargeDiff}), and is very useful for the design of new lubricants, since $E_\mathrm{adh}$ is usually more easily obtained by experiments than $\Delta V$.

As a second step we analyze how $\Delta V$ is related to the charge redistribution occurring during sliding. Redistribution of total charge~\cite{phillips:69} has been correlated before to stacking fault energies of II-VI and III-V compounds (which are related to the PES)~\cite{takeuchi:85}. However, the partition of total charge into subsystems is somewhat arbitrary~\cite{catlow:83} while our approach is based on the charge density which is unambiguously defined and can be evaluated using structure factors obtained from X-ray diffraction data with high accuracy~\cite{koritsanszky:01,kasai:18}.
As a general trend we observe that for not ideal stacking the charge density is lower in the interface region than for the minimum configuration. In Fig.\ref{fig:totCharge} we show this using charge density profiles for lateral configurations corresponding to the absolute minima and maxima of the PES for the same materials as in Fig.~\ref{fig:ChargeDiff}(a)-(c) as well as Cu(001). As can be seen in the insets, the charge density profile at the center of the interface is lower for the maximum configuration than for the minimum configuration for all systems.

We can thus envisage an experiment where the simultaneous measurement of $\Delta V$ and $\Delta \rho$ is performed and our finding is verified.
Such experiment is schematically represented in Fig.~\ref{fig:Corrugation}. The interaction energy between a probe moving on a crystalline substrate changes as a function of the relative lateral positions of the two bodies, as shown in Fig.~\ref{fig:Corrugation}(a) for copper on copper. We plot $\rho$ in a slice at the center of the interface for three different lateral configurations: the hollow [Fig.~6(b)], which is a minimum, an intermediate position [Fig.~6(c)], and the on-top configuration [Fig.~6(d)], which is a maximum. The arrow in Fig.~\ref{fig:Corrugation}(a) is visualizing the sliding path. We see that for the ideal fcc stacking the charge density is high and is distributed  rather uniformly [black solid curve of
Fig.~5(d)], while for the on-top configuration [Fig.~6(d)],
significantly less charge [red dashed curve of Fig.~5(d)] is
concentrated at the interface and it is less homogeneous
than Fig.~6(b).
This interfacial charge differences can be obtained, e.g., by a combination of Kelvin probe force microscopy (KPFM), scanning tunneling microscopy and atomic force microscopy (AFM)~\cite{mohn:12}, or improved frequency modulated AFM~\cite{albrecht:15}. KPFM is used to measure the work function $\phi$ locally~\cite{sadewasser:02, filleter:08}. Changes of $\phi$ are directly related to charge density differences\footnote{The connection is: $\Delta \phi =  - \frac{e}{\epsilon_0} \Delta p$, whith the change in the surface dipole $\Delta p$ depending on $ \rho_\mathrm{diff}$, $\Delta p\left(z\right) = \int z \rho_\mathrm{diff}\left(z\right) dz$~\cite{leung:03}} at the surface/interface~\cite{ling:13}. It has also been shown that the work function of a material is correlated with its adhesive friction~\cite{li:04}. For the example of the Cu(100) interface we calculate a work function difference of $\Delta \phi = \unit[-130]{meV}$ between two different surface stackings corresponding to Fig.~\ref{fig:Corrugation}(b) and Fig.~\ref{fig:Corrugation}(d).
The interaction potentials between the tip and the surface, which for Cu(100) are represented in Fig.~\ref{fig:Corrugation}(e), are also influenced by $\Delta \rho$ and can be measured by frequency shifts of an AFM~\cite{caffrey:15}.

In summary, we have shown that the interfacial charge density and its variation during sliding are the basic physical quantities, which determine adhesion, the PES, and thus the friction of a given atomically flat solid interface. The fact that the charge density is able to completely define the physical properties of a system is, of course, a long-known result of density functional theory~\cite{hohenberg:64}, but here we have shown how to deduce important figures of merit on the tribological properties of an interface from $\rho$ and $\rho_\mathrm{diff}$.

The linear relationships between $\rho_\mathrm{redist}$, which quantifies the charge redistribution when an interface is formed from separated surfaces, and $E_\mathrm{adh}$ is a very interesting result that is relevant for the general field of interface science, beyond the tribological context in which it has been presented here. We have shown that this simple linear scaling holds for three different types of bonding: weak, physical vdW interactions, stronger metallic bonding with rather uniformly distributed charge, and very strong directional covalent bonding. Simple power law scaling of the PES corrugation $\Delta V$ with $E_\mathrm{adh}$ has also been discovered where all investigated systems cluster around a single curve.

We have shown that the effectiveness of a certain class of friction reducing lubricant additives is to lower $\rho_\mathrm{redist}$, which in turn leads to a reduced adhesion $E_\mathrm{adh}$ and corrugation $\Delta V$.

We have found a higher interfacial charge density if the investigated system is in an energy minimum than in a maximum, and we connect this variation with the corrugation of the PES. We have suggested that such a connection can be experimentally observed by the simultaneous measurement of electronic and frictional properties during sliding.

\begin{acknowledgments}
This work was partially supported by Materials Design
at the Exascale (MaX) GA 676598 H2020 EINFRA-2015-
1, and the University of Modena and Reggio Emilia
through the Fondi di Ateneo per la Ricerca (FAR) 2016
project. M.~C.~Righi acknowledges support by the European Union through the MAX Centre of Excellence (Grant No. 676598). The authors thankfully acknowledge CINECA
(Consorzio Interuniversitario del Nord est Italiano Per il
Calcolo Automatico) for supercomputing resources
through the project Italian SuperComputing Resource
Allocation (ISCRA) B StressRx.  Several pictures in this Letter were created with the help of \mbox{XCrySDen}~\cite{xcrysden}.
\end{acknowledgments}

\end{document}